\documentclass[twocolumn]{aa}
\usepackage{graphicx}
\usepackage{txfonts}

\begin{document}
\title{Fast transition of type-B QPO in the black hole transient XTE J1817-330}
\author{
	K. Sriram\inst{1} \and 
	A. R. Rao\inst{2} \and 
	C. S. Choi\inst{1} 
}
\institute{
Korea Astronomy and Space Science Institute, 
Daejeon 305-348, Republic of Korea\\
\email{astrosriram@yahoo.co.in}
\and 
Tata Institute of Fundamental Research, Mumbai 400005, India
}

\date{}

\abstract
%Context
{The evolution of different types of quasi-periodic oscillations (QPOs) and the coupled  
radiative/physical changes in the accretion disk are still poorly understood. In a few black hole binaries
it was found that fast evolution of QPOs is associated with spectral variations. Such studies in other black hole
binaries are important to understand the QPO phenomenon.   
}
%Aims
{For the black hole transient XTE J1817-330, we study fast QPO transitions and accompanying spectral variations 
to investigate what causes the spectral variation during the QPO transition.
}
% Methods
{Roy et al. (2011) found QPOs in ten RXTE observations of XTE J1817-330. 
We found that, among the ten observations, only one observation shows erratic dips in its X-ray light curve.
The power density spectra and the corresponding energy spectra were extracted and analyzed for the
dip and non-dip sections of the light curve.
}
% Results
{
We found that type-B $\sim$6 Hz QPO changes into type-A QPO in a few tens of seconds along with a flux decrease.
This transient evolution is accompanied with a significant spectral variation.
}
%Conclusions
{ We report a transient QPO feature and accompanying spectral variation in XTE J1817-330. 
Based on our findings, we discuss the origin of fast evolution of QPOs and spectral variations.
}

\authorrunning{Sriram, Rao, Choi}
\titlerunning {Fast transition of QPOs in XTE J1817-330}

\maketitle

\section{Introduction}
\label{s:intro}
 The {\it RXTE} era of observations of black hole binaries (BHBs) has given a wealth of 
information on the radiative and dynamical astrophysical processes associated with the accretion disk
in these objects. 
The black hole binaries often go through an outburst during which the source spends in three different 
states viz. low hard (LH) or hard state, steep power-law 
(SPL) state/very high (VH) state or intermediate (IM) state and thermal dominated (TD)/high-soft state 
(McClintock \& Remillard 2004; Belloni 2010). The LH state is often observed during the raise and decay phase of the outburst
and it is characterized by a strong power-law component ($\Gamma$$\sim$1.5), strong band-limited noise, 
high radio flux and evidences to suggest that the inner disk front is truncated at a large radius (Done et al. 2007). 
The TD state is generally observed at the peak of the outburst associated with a strong disk component, 
weak band-limited noise, low or no radio flux and the accretion disk believed to reach the
last stable orbit. 
The SPL state is observed at the peak of the outburst in a few sources e.g. XTE J1550-564, characterized by a steep power-law component, 
band limited noise with low and high frequency quasi-periodic oscillations (QPOs), jet is assumed to switch on/off in this state (Fender et al. 2004) 
and disk is assumed to be truncated at a small radius (Done et al. 2007). This state spans relatively more area in
the  hardness intensity diagram (HID) plane (Remillard \& McClintock 2006) and 
is further classified as hard and soft intermediate states (HIMS and SIMS) (Belloni 2010; Belloni et al. 2011).

The generating site and mechanism for the low frequency quasi-periodic oscillations (LFQPOs)
in black hole binaries/transients are still not properly understood. The possible regions in the accretion disk responsible for the QPO modulation
are thought to be the inner accretion disk or the Compton cloud in the hot inner region
(Done et al. 2007).
Casella et al. (2005) summarized the various types of LFQPOs seen in black hole 
binaries and provided the basis for the classification in three broad categories viz. type-A, B and C,
primarily depending on the rms amplitude and the quality factor ($\nu / \delta \nu$).
The properties and occurrence of the various types of QPOs are well studied using the hardness intensity diagram (HID).

 The type-C QPOs have variable centroid peak positions in the range of 0.1 -- 15 Hz 
along with a harmonic. They are  narrow ($\nu\ / \delta \nu$ = 7 -- 12) and have strong
variability strength  with a maximum fractional rms variability of $\sim$30\%.
Spectrally they are often associated with the LH state or radio loud HIMS state. Recently in H1743-322, Miller-Jones et al. (2012) observed that
the disappearance of type-C QPO is associated with a jet ejection event.
 Type-B QPOs are observed at $\sim$6 Hz, relatively weaker (rms $\sim$4\%) and are moderately narrow ($\nu\ / \delta \nu$ 
$\sim$  6). 
Some times a weak harmonic is also observed and interestingly they are sometimes accompanied by a rapid transition on a timescale of a few seconds (Nespoli et al. 2003). 
Type-B QPOs are observed as the source traverses from the hard intermediate state (HIMS) to the soft intermediate state (SIMS). 
A few studies also show that the type-B QPO is connected to the emission and collimation of radio jets (Casella et al. 2004; Fender et al. 2004). 
The type-A QPO is weaker and broader than type-B QPO with a centroid frequency at $\sim$8 Hz and have similar spectral properties as of type-B QPO's spectral properties
(Belloni et al. 2011; Belloni 2010; Homan et al. 2005).  

Transitions among different types of QPOs have been observed in different black hole sources. In GX 339-4
Nespoli et al. (2003) found a transient QPO at 6 Hz and related it with a spectral hardening during the HIMS to SIMS 
transition. Casella et al. (2004) observed a fast transition in the properties of Power Density Spectra (PDS) and the
corresponding energy spectra in the black hole transient XTE J1859+226. 
Soleri et al. (2008) found  type-B QPOs in GRS 1915+105 
along with the transitions from type-B to A and B to type-C as the source moves from the HIMS to SIMS and vice-versa. 
Recently Stiele et al. (2011) reported that during the soft to hard state transition in SIMS (GX 339-4), the low 
frequency of type-B QPO is characterized by a low power-law index when compared to the hard to soft state transition. 
These studies suggest that the associated physical changes in accretion disk are complex and could be related to the onset
of radio jets, which is often associated with the HIMS/SIMS transition (the so-called `jet-line') (Fender et al. 2004; Migliari \& Fender 2006; Fender et al. 2009). 

The source XTE J1817-330, an X-ray transient, was discovered on 2006 January 26 with the All-Sky Monitor onboard 
the {\it RXTE} satellite. The observations in radio (Rupen et al. 2006), optical (Torres et al. 2006),
and near infrared/ultraviolet (D' Avanzo et al. 2006; Sala et al. 2007) wavelengths suggest that it is a strong candidate 
for black hole binary (BHB) system. The spectral studies show that during the peak of the outburst, the source was in the 
high-soft (HS) state (see Remillard \& McClinktock 2006 for spectral classification) and later went to the
low-hard state as the source intensity gradually decreases (Gierlinski et al. 2008; Roy et al. 2011). Only in ten observations 
QPOs were observed with a limited frequency domain of 4-6 Hz (Roy et al. 2011) and in one occasion QPO was observed at 8--9 Hz 
(Homan et al. 2006; Roy et al. 2011). It was reported that all the QPOs were found in the HS state, 
though in the HID plane the corresponding observations could be associated to the HIMS/SIMS region (Roy et al. 2011). 
The observed HS state in XTE J1817-330 has a unique signature, a high disk temperature($\sim$0.8-0.9 keV) along with 
a relatively hard spectral component ($\Gamma$$\sim$2.1-2.3), which is not often seen in other BHBs.

In the present study, we report a transient QPO feature in XTE J1817-330 which shows a sudden transition from type-B to type-A QPO. 
The spectral change during the observed QPO transition is discussed.
                   
\section{Data Reduction and Analysis}
\label{s:Reduction and Analysis}
This study uses the archival data obtained by the {\it RXTE} satellite (Bradt et al. 1993). There are 160 observations for 
XTE J1817-330 among which ten observations show QPOs in their PDS (Roy et al. 2011). We used
the 16 s bin data from the Proportional Counter Array (PCA, Jahoda et al. 2006) to study
intensity variations in the ten observations. We found that only one observation, ObsId 91110-02-30-00 (MJD 53790), 
shows erratic dips in the light curve. In Figure 1 we show the RXTE PCA count rates during the outburst. The
observation showing the dips  lies on the decay phase of the outburst when
a flare-like feature was observed in the higher energy bands (Fig. 1) and was not seen in the lower energy bands (Roy et al. 2011).
The PDS was extracted using the single bit mode data SB\_125us\_8-13\_1s (3.68-5.71 keV) and 
SB\_125us\_14-35\_1s (6.12-14.76 keV) with a time resolution of 1/1024 s. We have investigated the
PDS behaviour separately for the 
dip and non-dip section of the light curve and the details of the observations are given in  Table 1. 
We extracted the energy spectra for the respective sections from PCU2 unit (3-25 keV) and added a systematic error of 0.5\%.  
We also extracted HEXTE cluster B (High-Energy X-ray Timing Experiment, Rothschild et al. 1995) spectra (15-100 keV)
excluding the detector 2 for dip and non-dip sections. 
The source and background data were obtained using the hxtback command of FTOOLS and extracted the corresponding spectra. 
The dead-time correction was applied to the obtained spectra using hxtdead command (see e.g. Sriram et al. 2007). 
We used HEASOFT v6.8 for the data filtering and used spectral models available in XSPEC v12.5.0 (Arnaud 1996).

\subsection{Temporal Analysis}
The X-ray light curve for ObsId 91110-02-30-00 is shown in
Figure 2 (top panel) for the 2 -- 20 keV energy band. A series of dips,
lasting for 10 - 100 s, can be seen in the light curve. For detailed investigations
we have segregated the data into non-dip and dip sections. Samples are indicated as rectangular boxes in Figure 1.
The PDSs for the non-dip and dip sections in the high (6.12-14.76 keV) 
and low energy band (3.68-5.71 keV) are shown in the middle and bottom panels of 
Figure 2 along with the best-fit models consisting of a power-law and a Lorentzian.
The PDS in higher energy band of the non-dip section clearly shows a type-B QPO which has a centroid frequency of 5.57 $\pm$ 0.06 Hz  and
quality factor (Q=$\nu$/FWHM) of 5.74 and rms of 12 $\pm$ 1\%. The PDS of the dip section shows a type-A QPO with 
a centroid frequency of 6.00 $\pm$ 0.73 Hz and quality factor Q $\le$ 1 with rms of 14 $\pm$ 3\%. 
The important signature of type-B QPO is its harmonics which are 
seen in the PDS of the non-dip section and they are absent in the type-A QPO (dip section). 
We have examined the energy dependency of the QPOs and found that the type-B QPO seen in
the non-dip section has the same frequency (5.54 $\pm$ 0.07 Hz) and quality factor (5.32) in the lower energy band 
(3.68 - 5.71 keV), but the rms value is lower (6.1 $\pm$ 0.5\%). Interestingly, in the lower energy band, 
the dip section PDS shows a QPO at 2.67 $\pm$ 0.28 Hz with a Q $\le$ 1 and rms of 7 $\pm$ 2\%.
%In the low energy band, the dip section QPO (type-A) shows very little power (XX\% compared to YY\%).
%PDS show a QPO at 2.67 Hz with a Q $\le$1. 
It is clear that during the dip section, the centriod frequency of QPO has significantly changed in the energy domain.

We have  examined the dynamical PDS of the light curve and found that in the
non-dip section the centroid of the QPO often varied from 5.0 to 6.5 Hz.

\subsection{Spectral Analysis}

To see whether there is a significant spectral difference between the non-dip and dip sections, we first checked the
residuals by fixing the non-dip spectral parameters to the dip spectrum.
The residuals are shown in the top panel of Figure 3 and the high $\Delta \chi$ values of the dip spectrum suggest that
the spectrum has changed significantly during the type-B to A transition.
We then fitted the non-dip and dip spectra (3-25 keV) using the model {\it wabs(diskbb+Gaussian+power-law)} % hereafter model 1,
where the hydrogen equivalent column density was fixed at N$_{H}$ (wabs model)= 1.2$\times$10$^{21}$ cm$^{-2}$ (Rykoff et al. 2007)
and the Gaussian line centroid energy as 6.4 keV (for iron line). 
The best-fit spectral parameters are shown in Table 2 and the unfolded spectra are shown in the bottom panels of Figure 3.
We found that the power-law index has decreased from $\Gamma$=2.46$\pm$0.03 (non-dip) to $\Gamma$=2.29$\pm$0.03 
(dip), indicating that a spectral hardening occurs during the type-B to A transition. 
We also fitted the PCA+HEXTE spectrum (3-100 keV) using the same model and observed similar hardening of power-law index. 
 Both PCA and PCA+HEXTE spectral fits result shows that the power-law flux decreased from non-dip to dip section 
and the flux is decreased by $\sim$17\% in the energy band 3--100 keV, whereas no appreciable 
change is observed in the diskbb model component. The total flux obtained by Roy et al. (2011) (i.e. 1.02 $\times$ 10$^{-8}$ erg cm$^{-2}$ s$^{-1}$ in 3-25 keV) 
and in the present work (1.04 $\times$ 10$^{-8}$ erg cm$^{-2}$ s$^{-1}$ in 3-25 keV) are similar.  
We note that the results do not vary if we free the hydrogen equivalent column density and Gaussian line energy. 
An attempt was made to use the CompTT model instead of the power-law, but it resulted in unacceptable fits to the spectra. 
The spectral hardening along with the flux decrease in the dip suggest that a relatively cooler portion of 
hard X-ray emitting region is ejected away leaving behind a hotter hard X-ray emitting region. 

\section{Discussion and Conclusion}
\label{s:summary}

The low frequency QPOs (LFQPOs) are the most common temporal features observed 
in BHBs and they play a key role in the spectral classification schemes 
(Remillard \& McClintock 2006; Belloni 2010; Belloni et al. 2011). 
During the outburst, the frequency of LFQPO increases 
and it disappears at HS state and again re-appears in the VHS/IM state. However, the 
exact production mechanism of oscillation and the location in the accretion 
disk is not yet clear but it is thought to occur in the inner regions of the accretion disk. 
The classification of the type-C, B and A QPOs 
(Casella et al. 2004, 2005) shows that there is a unique but complex connection 
with the soft and hard X-ray components and indicates toward a sudden state 
transition from HIMS to SIMS or vice-versa (Belloni et al. 2005; Belloni 2010). 
In general type-B, A QPOs and their transitions are often observed in BHBs when they  are in very 
high (VH)/steep power-law (SPL) state or intermediate state (IM) (Belloni 2010). 
In this state the accretion disk is considered to contain a compact corona (low temperature with high optical depth)
along with a high temperature Keplerian disk located close to the last stable orbit (Done \& Kubota 2006; 
Sriram et al. 2007, 2009, 2010).

We found one such observation in XTE J1817-330, where a type-B QPO of $\sim$ 6 Hz is transiting to a type-A QPO 
in a few tens of seconds. During this observation, the source was in the declining phase of the outburst 
showing a sudden flaring in the higher energy bands (Fig. 1). It is evident from the PDS that the characteristic 
harmonics have also disappeared during the transition (middle panel of Fig. 2) and the energy dependent PDS 
study of dip section shows that the centroid frequency of QPO is shifted to 
lower frequency from higher to lower energy band, whereas no such variation was seen in non-dip section.
This could be due to some ejection of low temperature material from the accretion disk. The spectral results during the type-B and type-A QPO
show that the power-law component became harder along with a decrease in its flux (Table 2). 
%Similar variations were observed when only the HEXTE spectra are used, confirming this variation. 
This observation falls in HS classification but occupies SIMS location in the HID plane (Roy et al. 2011). 
This source has a unique property that in its HS state, the hard component
is relatively stronger with the occasional presence of QPOs which is generally not observed in other BH sources. 
However in the transient QPO observation, we suggest the dips may occur from a sudden transition from the soft to SIMS state,
which is inferred from the variation in the power-law index.      
 
The transition of QPOs among different types viz. type C-B and type B-A and 
vice-versa were observed in a few other black hole sources. 
Homan et al. (2001) found broad QPO features around 6 Hz in XTE J1550-564 when the source was in a low intensity
state and it became narrow in a high intensity state. In case of XTE J1859+226, a broad 6 Hz feature was observed
when the source was in a high intensity state and became narrow in a low intensity state (Casella et al. 2004).
In GX 339-4, 
Nespoli et al. (2003) observed a transition of a type-A 6 Hz QPO (low intensity state) to a type-B along with a harmonic
associated with the hardening of the spectrum. 
They found that during the transition, the disk flux decreased by 9\%, power-law
component flux increased by 30\% and total flux increased by 9\%. Belloni et al. (2005) found that
this transient feature is associated with the spectral transition from HIMS to SIMS.  
Similar transient variation (type-B to A QPO) was also observed in H1743-322
(Homan et al. 2005, see their Figure 4 panel C and D), however no appreciable changes were observed
in disk and power-law components but during this transition {\it Chandra} observations revealed 
absence of Fe lines possibly caused by the high ionization or by absorbers (Miller et al. 2006).
GRS 1915+105 also shows such transient type A-B 
QPO features and their properties are dependent on the location in HID and were discussed in the context of 
onset of a variable jet (Soleri et al. 2008). Recently Stiele et al. (2011) found that in case of GX 339-4, 
the power-law index strengthens as the source moves from the soft to hard state in SIMS and weakens as the source moves from 
the hard to soft state.  

The spectral result shows the spectral hardening 
of power-law component in dip (Table 2), suggesting that the source was 
traversing from the soft to hard SIMS. The overall power-law component 
variation can be explained in the following scenario. Assume a Compton 
cloud with a radial distribution of electron temperature
i.e., hotter at the center and cooler at the outer region of the Compton cloud. 
From this configuration, as the source moves
to the low intensity state (i.e., dips) the outer region of Compton cloud 
is ejected away in the form of a jet or an outflow from the disk and leaves
behind a hotter part of the Compton cloud. This physical picture explains 
both the spectral hardening of power-law component and the flux decrease in the dip.
 Moreover this scenario is also favored by 
the shifting of the centroid frequency of QPO (low energy band) in the dip section 
which suggests that the low temperature component is relatively more effected than 
the high temperature component.

Type-B QPOs are relatively rare and they tend to occur during the local peaks of the outbursts
(see Motta et al. 2011). Such fast transitions from type-B to type-A are still more rare.
In GX 339-4 such `flip-flop' transitions were seen in the {\it Ginga} data 
(Miyamoto et al. 1991) and later in the {\it RXTE} data (Motta et al. 2011). Takizawa et al. (1997)
reported `flip-flop' transitions in GS 1124-683 and Casella et al. (2004) reported it in
the {\it RXTE} data of XTE J1859+226. The fast transitions seen in GRS 1915+105
in the $\mu$ and $\beta$ classes are also of similar nature (Soleri et al. 2008).
As discussed by Miyamoto et al. (1991), a jet origin for the origin of some part
of the hard component (and possibly for the type-B QPOs) is a feasible scenario.

In summary, we found a transient 6 Hz QPO in the source XTE J1817-330 during the decay phase of the outburst. 
During this evolution the type-B QPO characteristics in non-dip of the light curve are changing into type-A QPO in dip.
Our spectral study substantially provide one of the strong evidences for spectral variations during the fast transition of type-B to type-A QPO.
This reported result for the source XTE J1817-330 again give credence to the physical scenario
that the type-B to A QPO transition occurs in a short timescale and could be connected to
variable onset of jet or an outflow in the accretion disk.

\begin{acknowledgements}
%\acknowledgements 
We thank the anonymous referee for the very useful comments. 
This research has made use of data obtained through the HEASARC Online Service, 
provided by NASA/GSFC in support of the NASA High Energy Astrophysics Programs. 
KS thanks the hospitality provided at TIFR, where some part of the work was done. 
\end{acknowledgements}
%%%%%%%%%%%%%%%%%%%%%%%%%%%%%%%%%%%%%%%%%%%%%%%%%%%%%%%%%%%%%%%%%%%%%%%%%%%%%%%%%
\clearpage

\begin{table}
\begin{minipage}[t]{\columnwidth}
\caption{The details of the observation analyzed in the present work for the source XTE J1817-330.} 
\label{tab1}
\centering
\renewcommand{\footnoterule}{}
\begin{tabular}{cccccc}
\hline
\hline
ObsId& MJD (Date)& \multicolumn{3}{c} {Duration (s)} & \\
\hline
&&Total observation &Non-dip&Dip\\
\hline
91110-02-30-00&53790 (2006-02-24)&4132 &990 &350\\
\hline
\hline
\end{tabular}
\end{minipage}
\end{table}

\begin{table}
\begin{minipage}[t]{\columnwidth}
\caption{Best-fit spectral parameters for the non-dip and dip spectra using the model 
wabs(diskbb+Gaussian+power-law) (see text). 
The quoted errors are at a 90\% confidence level and fluxes are in 
units of 10$^{-9}$ ergs cm$^{-2}$ s$^{-1}$ for PCA (3-25 keV) and 
HEXTE (15-100 keV). The fluxes for PCA+HEXTE spectral fits are obtained 
in the energy band 3--100 keV.} 
\label{tab1}
\centering
\renewcommand{\footnoterule}{}
\begin{tabular}{ccc}
\hline
\hline
Parameters&\multicolumn{2}{c}{ObsId 91110-02-30-00} \\
\hline
&Non-dip (type-B QPO)&Dip (type-A QPO)\\

\hline
\hline
&For PCA&\\
\hline
\hline

%N$_{H}$($\times$ 10$^{22}$ cm$^{-2}$)\footnote{Hydrogen equivalent column density} &$1.4\pm0.3$ & 3.2$\pm$0.5\\

$kT_{in}$ (keV)\footnote{Inner disk temperature.}&$0.90\pm0.01$ &0.91$\pm$0.01\\
$N_{diskbb}$\footnote{Normalization of diskbb model.}&1289$\pm$115&1411$\pm$150\\

$\Gamma$\footnote{Power-law index.}&2.46$\pm$0.03& 2.29$\pm$0.03  \\
$\Gamma_{Norm}$\footnote{Normalization of the power-law model (photons keV$^{-1}$cm$^{-2}$s$^{-1}$ at 1 keV).}&4.59$\pm$0.37 & 2.74$\pm$0.45 \\
diskbb flux&4.37& 4.41\\	
power-law flux&6.10&5.00 \\
Total flux&10.43&9.41 \\
$\chi^{2}$/dof&48/48 &37/48 \\
\hline
\hline

&For PCA+HEXTE &\\
\hline
\hline
$kT_{in}$ (keV)&0.92$\pm$0.008 &0.91$\pm$0.01\\
$N_{diskbb}$&1271$\pm$66&1413$\pm$155\\

$\Gamma$&2.46$\pm$0.02& 2.31$\pm$0.04  \\
$\Gamma_{Norm}$&4.71$\pm$0.37 & 2.89$\pm$0.42 \\

diskbb flux&4.40& 4.45\\	
power-law flux&7.20&6.14 \\

$\chi^{2}$/dof&104/79 &71/79 \\

\hline
\hline
\end{tabular}
\end{minipage}
\end{table}

\clearpage
%%%%%%%%%%%%%%%%%%%%%%%%%%%%%%%%%%%%%%%%%%%%%%%%%%%%%%%%%%%%%%%%%%%%%%%%%
\begin{figure}
%\centering
%\resizebox{\hsize}{!}
{\includegraphics[height=18cm,width=18cm,angle=0,clip=]{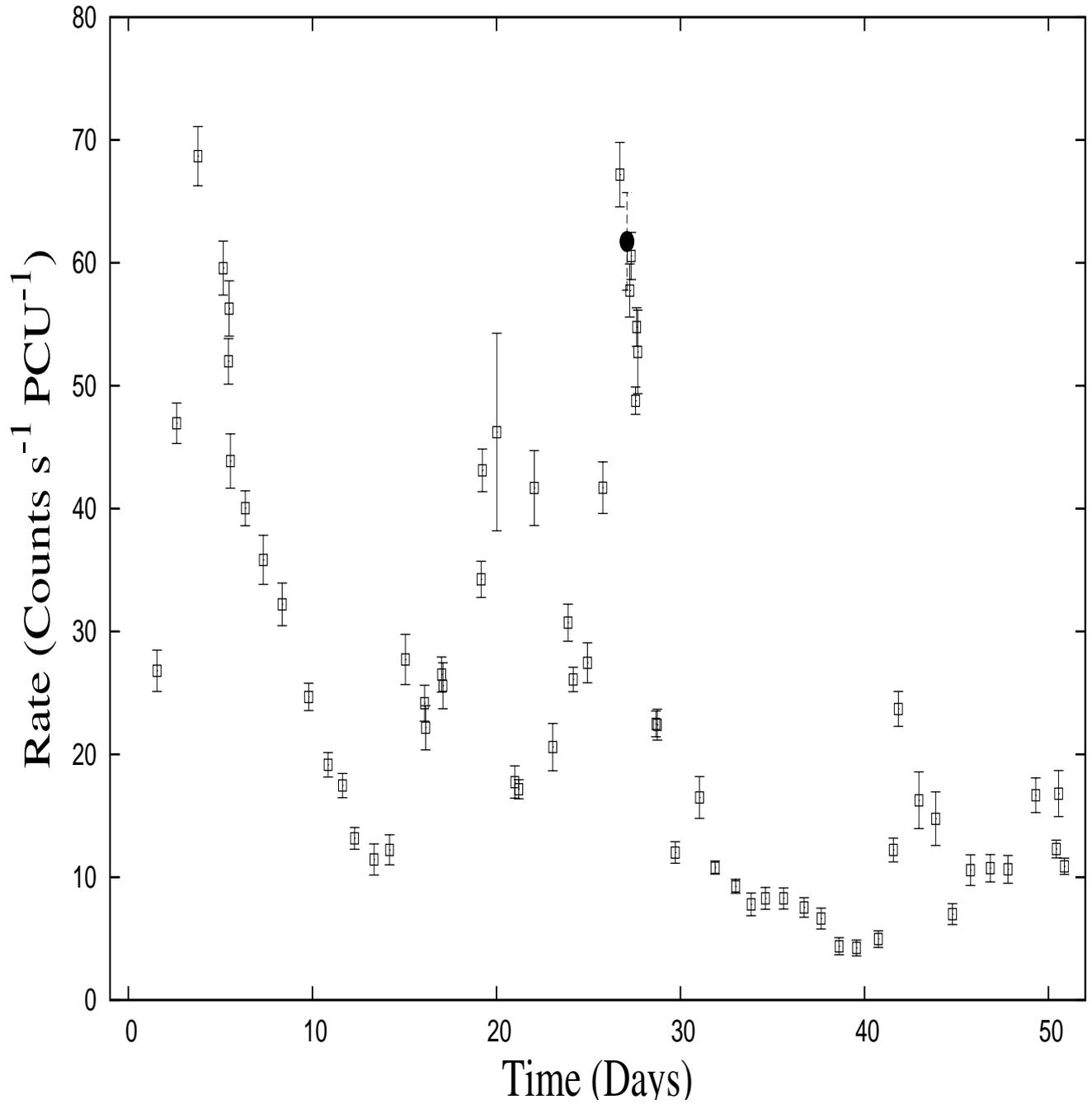}}\\%for printer format for aa
%{\includegraphics[height=18cm,width=10cm,angle=-90,clip=]{qpo_all.ps}}%for printer format for aa
\caption{The 9-20 keV PCA light curve of the outburst for initial 50 days (53764--53814) is shown. 
The filled circle marks the observation (MJD 53790) during which the transient QPO is observed.} 
       %\label{Fig1}
 \end{figure}

\clearpage

\begin{figure}
%\centering
%\resizebox{\hsize}{!}
{\includegraphics[height=18cm,width=7.5cm,angle=-90,clip=]{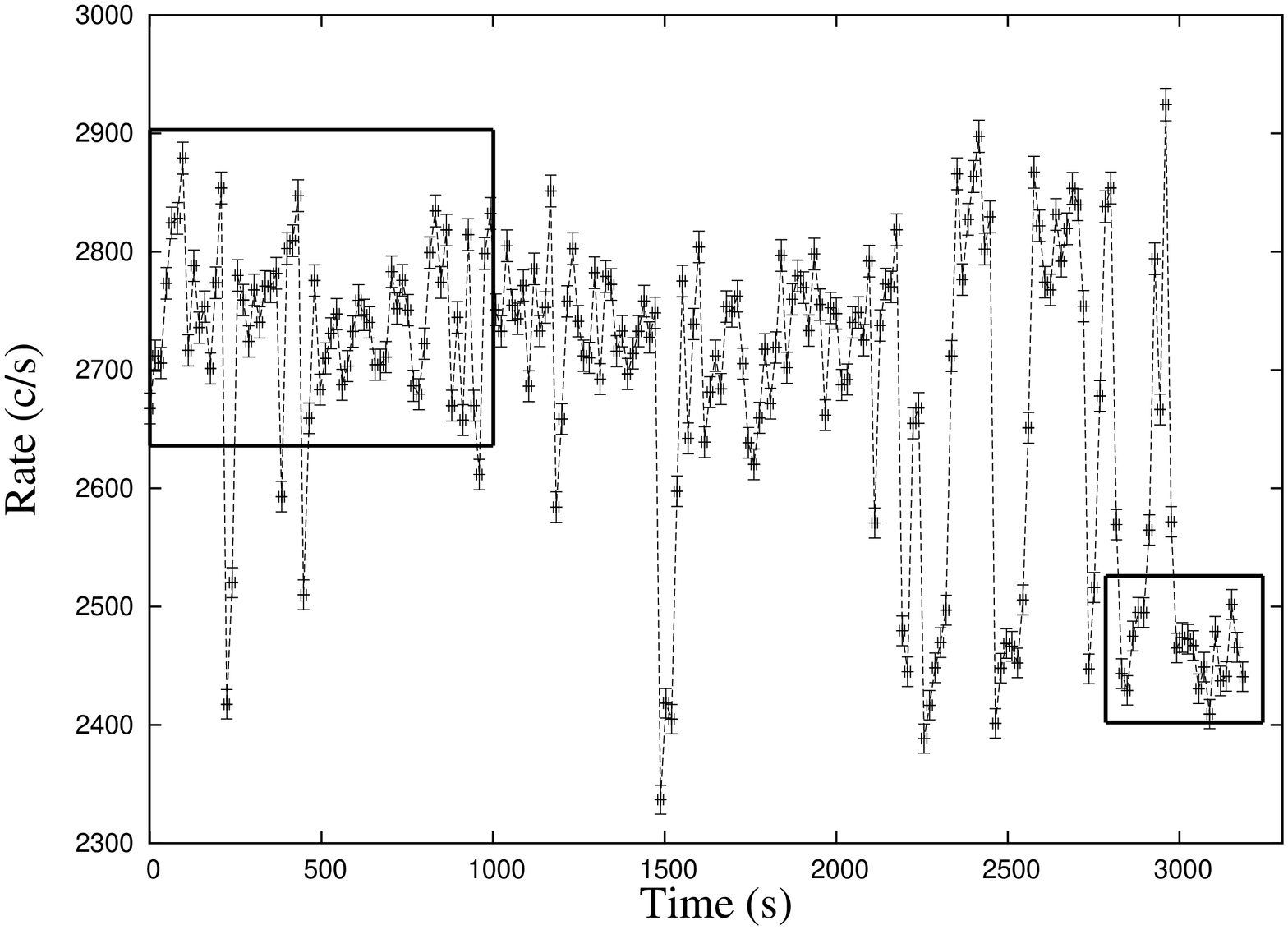}}\\%for printer format for aa
{\includegraphics[height=18cm,width=7.5cm,angle=-90,clip=]{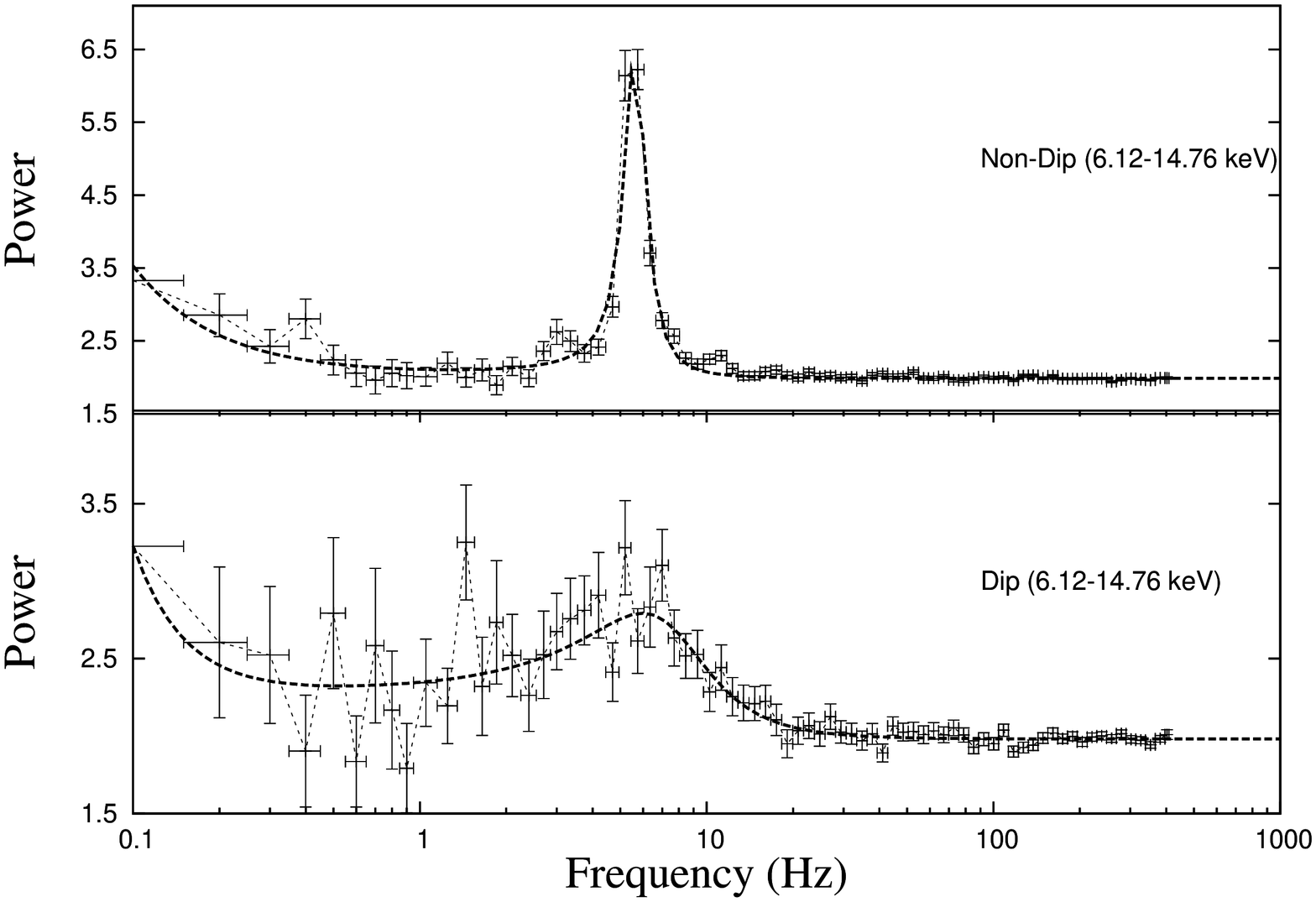}}\\%for printer format for aa
{\includegraphics[height=18cm,width=7.5cm,angle=-90,clip=]{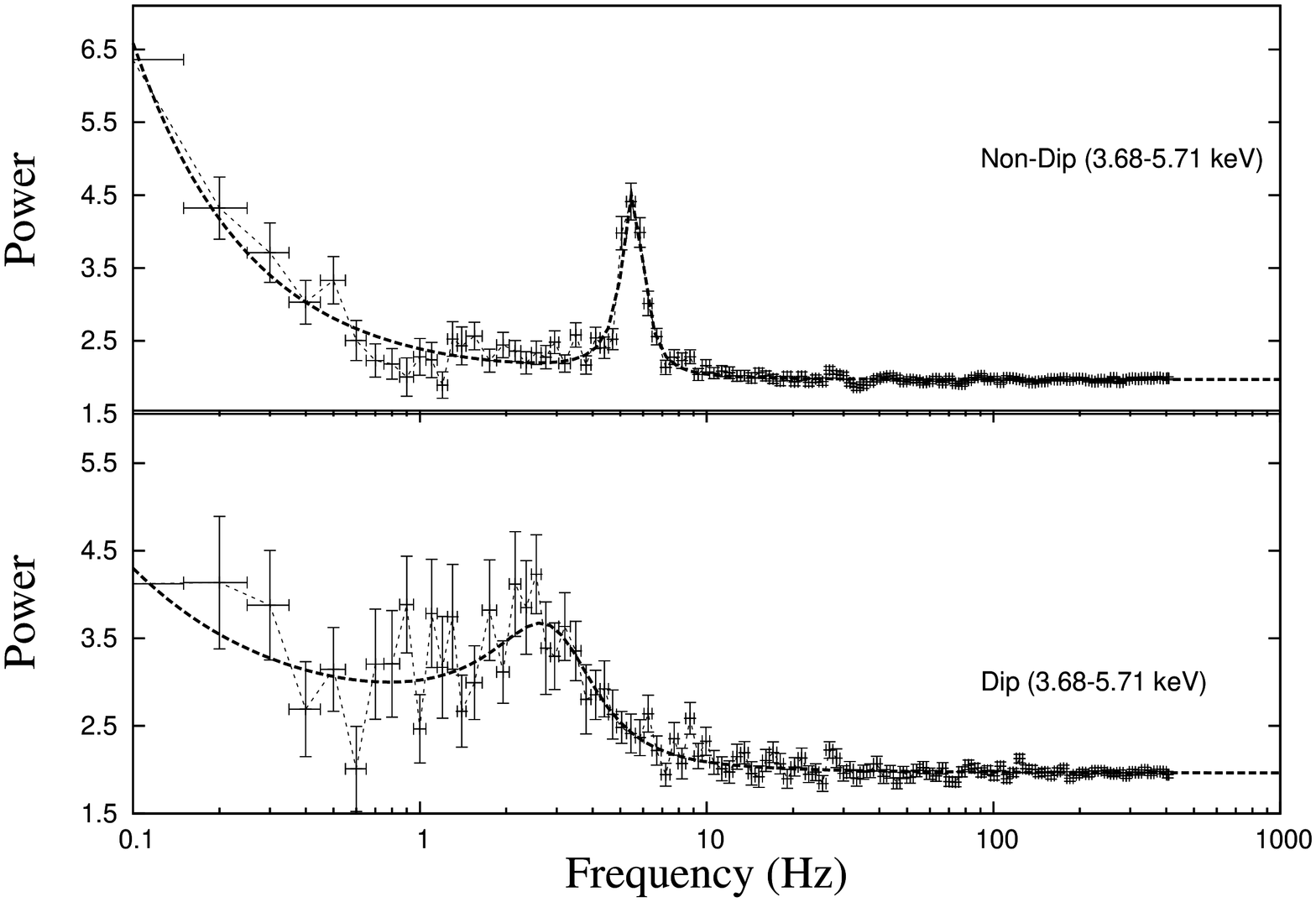}}
\caption{Top: The PCA 16 s bin light curve in the 2-20 keV band. The boxes show the sections 
of non-dip and dip used for PDS and spectral analysis. Middle: 
{\bf The PDS of non-dip and dip sections in the energy band 6.12--14.75 keV along with their best-fits using   
a {\it power-law+Lorentzian+constant} model (shown in thick dashed lines). The wings or harmonics in the non-dip PDS are clearly 
visible but absent in the dip PDS, which is a characteristic feature 
observed when a type-B QPO moves into a type-A QPO. Bottom: Similar PDSs in the lower energy band (3.68--5.71 keV) are shown.}
The Poisson noise is not subtracted. } 
       %\label{Fig1}
 \end{figure}
\clearpage

\begin{figure}
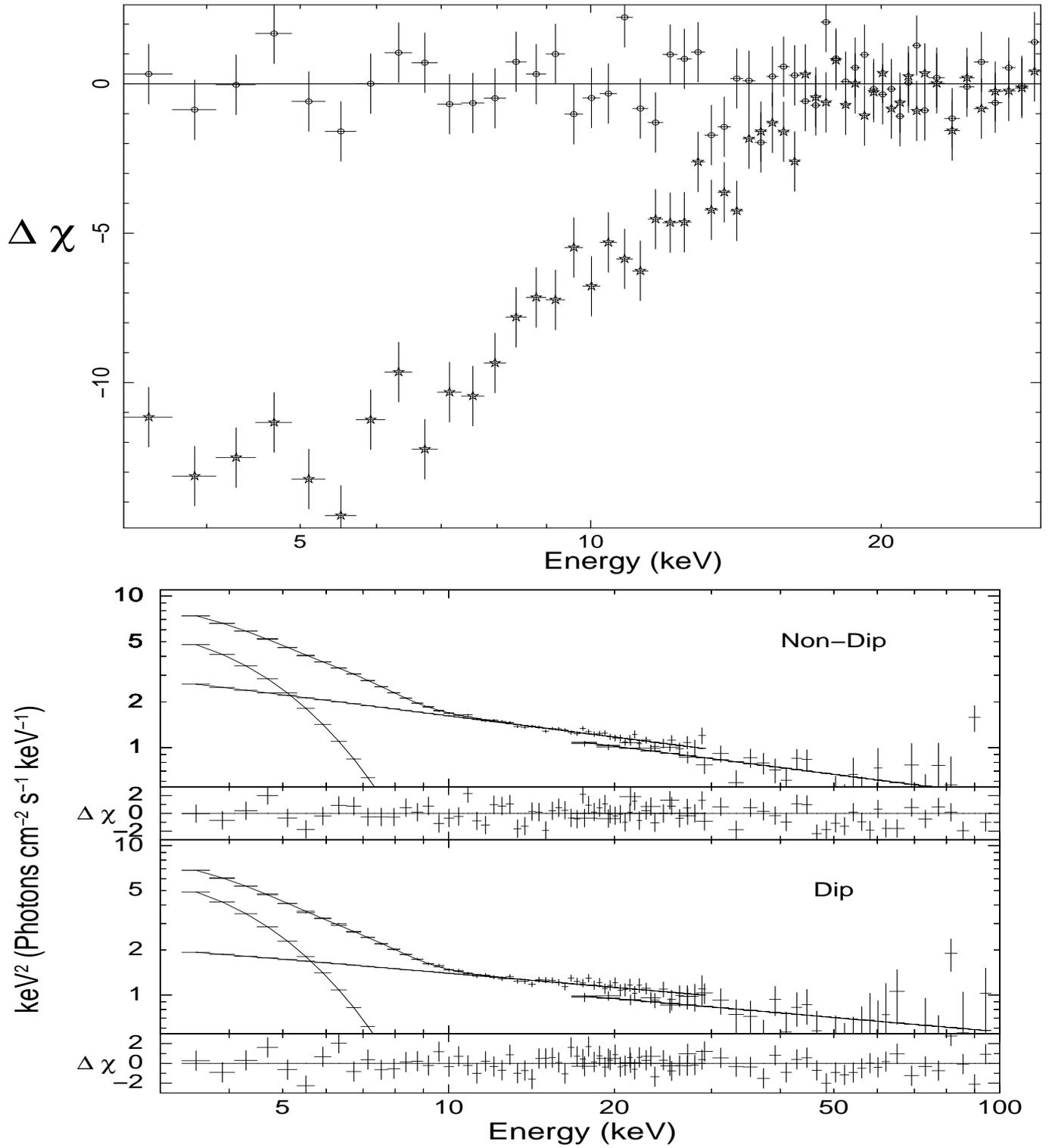

%\centering
%\resizebox{\hsize}{!}
{\includegraphics[height=18cm,width=10cm,angle=-90,clip=]{fig3.1.ps}}\\%for printer format for aa
{\includegraphics[height=18cm,width=10cm,angle=-90,clip=]{fig3.2.ps}}\\%for printer format for aa
     \caption{Top: The residuals of the non-dip and dip spectra obtained by applying
the best-fit spectral parameters of the non-dip spectrum. 
The high {\bf residuals} of the dip spectrum (star symbols) show that the spectrum has
appreciably changed which is responsible for the observed type-A QPO in this section.
Bottom: The spectral fit to the broadband spectra (3-100 keV) of  the non-dip and dip sections.
}
       %\label{Fig2}
 \end{figure}

%%%%%%%%%%%%%%%%%%%%%%%%%%%%%%%%%%%%%%%%%%%%%%%%%%%%%%%%%%%%%%%%%%%%%%%%%%%%%%%%%%%%%%%%

%%%%%%%%%%%%%%%%%%%%%%%%%%%%
\end{document}